\documentclass[useAMS,referee]{biom}

\def\bSig\mathbf{\Sigma}

\let\bibhang\relax
\usepackage{natbib}
\usepackage{booktabs, multirow, tabularx, graphicx,caption}
\usepackage{amsmath}
\usepackage{bbm}

\newtheorem{condition}{Working condition}

\title[Data integration for estimating subgroup-specific CATEs using coarsened external information in RCTs]{Data Integration for Estimating Subgroup-Specific Conditional Average Treatment Effects (CATEs) Using Coarsened External Information in Randomized Trials}

\author{Youqi Yang$^{1,*}$\email{youqi@umich.edu}, 
Walter Dempsey$^{1,2}$, and Bhramar Mukherjee$^{3,4,5}$ \\
$^{1}$Department of Biostatistics, University of Michigan, Ann Arbor, Michigan, U.S.A. \\
$^{2}$Institute for Social Research, University of Michigan, Ann Arbor, Michigan, U.S.A. \\
$^{3}$Department of Biostatistics, Yale University, New Haven, Connecticut, U.S.A. \\
$^{4}$Department of Epidemiology (Chronic Diseases), Yale University, New Haven, Connecticut, U.S.A. \\
$^{5}$Department of Statistics and Data Science, Yale University, New Haven, Connecticut, U.S.A. \\}

\begin{document}

\pubyear{2025}
\artmonth{July} 

\pagerange{}

\doi{} 

\providecommand{\newblock}{}

\label{firstpage}

\begin{abstract}
Randomized controlled trials (RCTs) are often underpowered to detect treatment heterogeneity in subgroups defined by cross-classifications of multiple covariates, due to sparse sample sizes in some strata. External RCT data can help, but typically provide treatment effect estimates at a coarser level (e.g., by sex or race) rather than for the finer subgroups of interest (e.g., race-by-sex). We propose a novel James–Stein (JS)–type estimator that borrows strength from such coarsened external estimates to improve estimation of finer subgroup-specific conditional average treatment effects (CATEs) in an internal study, while accommodating potential incompatibility in marginal CATEs across populations. Based on asymptotic theory, we derive a practical analytic variance estimator for the JS estimator that exhibits acceptable empirical performance. Under mild conditions, we show that the proposed estimator uniformly dominates the ordinary least squares (OLS) estimator based on internal data regarding a weighted quadratic loss. Simulation studies demonstrate favorable performance compared with existing shrinkage methods, including empirical Bayes and generalized ridge estimators. We illustrate our method by estimating race-by-sex subgroup CATEs in a tirzepatide weight-loss trial (SURMOUNT-1), borrowing sex-specific and race-specific estimates from two previous semaglutide trials (STEP 1 and STEP 2). The proposed method detects a significantly larger treatment effect on percentage weight loss in the female-White subgroup than in the female–Asian subgroup, a difference not detected using internal data alone.
\end{abstract}


\maketitle

\section{Introduction}

There is growing awareness that average treatment effects (ATEs) reported in randomized controlled trials (RCTs) can mask important variations in how individuals or subgroups of patients respond to treatment \citep{kent2010assessing}. Pre-specified subgroup analyses are a common tool to explore these differences, often considering pre-treatment covariates such as sex or race. However, these stratified analyses are typically powered to detect marginal effects within univariate subgroups and often lack sufficient sample size to uncover more granular treatment effect heterogeneity \citep{wang2007statistics}, particularly when treatment effects vary across combinations of patient characteristics (e.g., race-by-sex interaction). 

Improving power to detect such heterogeneity may therefore require borrowing information from external RCTs. In practice, however, external evidence is often available only in aggregate form, most commonly as point estimates and confidence intervals (CIs), rather than as individual-level data \citep{tierney2020comparison}. These summaries typically estimate ATEs or univariate subgroup-specific conditional average treatment effects (CATEs), which may be defined at a coarser level of granularity than the CATEs of interest in a new study. Moreover, even at this coarser level, external causal effects may not be fully compatible with their internal counterparts. To address these challenges, we propose a novel James–Stein (JS)–type method for estimating finely stratified CATEs that incorporates potentially incompatible coarsened external information.

\begin{figure}
    \centering
    \includegraphics[width=1\linewidth]{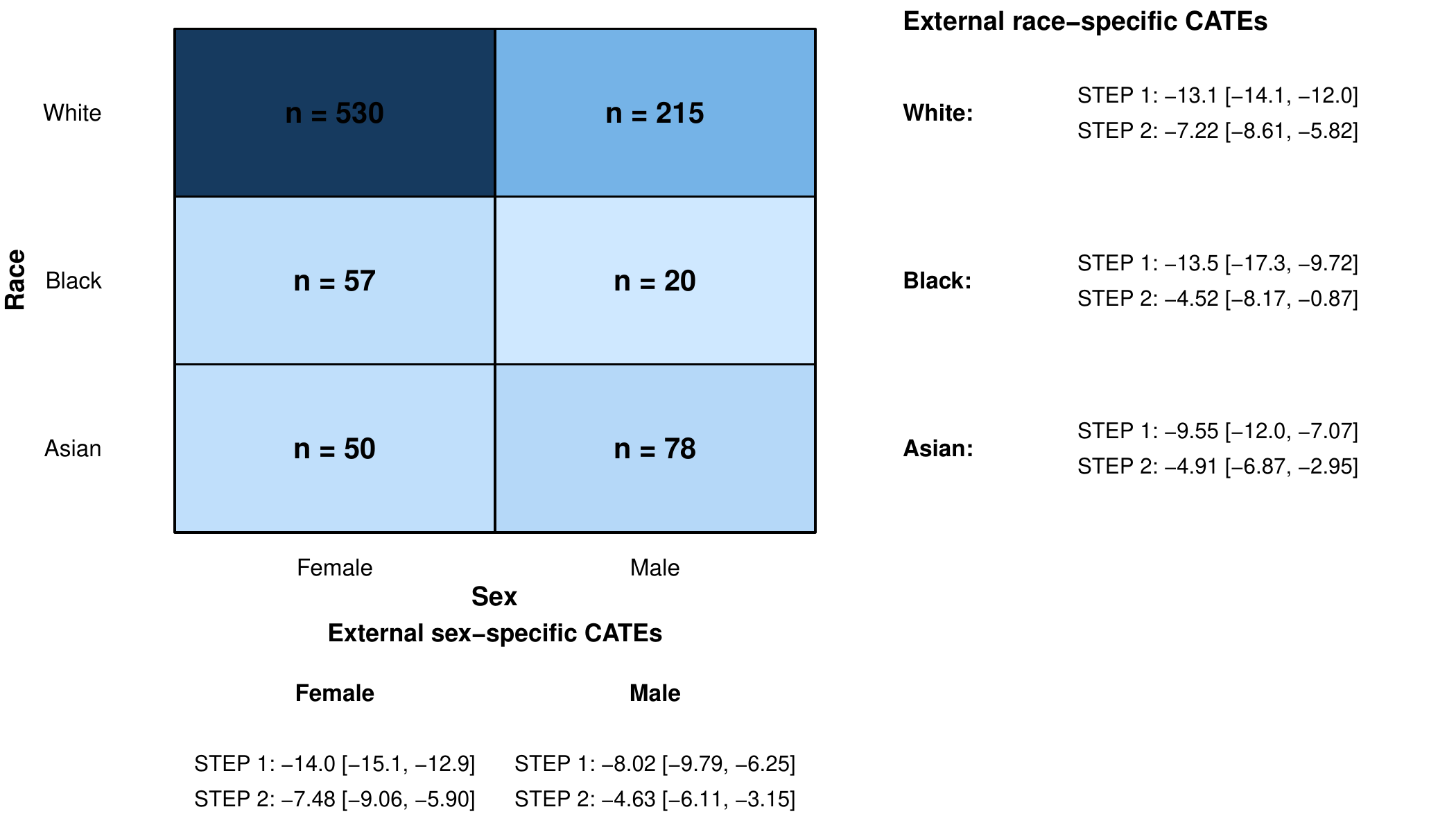}
     \caption{\textbf{Race-by-sex subgroup sample sizes from SURMOUNT-1 and external summary information from STEP~1 and STEP~2.} Each box shows the number of participants in each race-by-sex subgroup in SURMOUNT-1 among individuals with complete data; darker shading indicates larger sample sizes. External information from STEP~1 and STEP~2 includes published CATE estimates by race and by sex for the percentage change in body weight from baseline. Estimates are reported as means with 95\% confidence intervals for the comparison of semaglutide 2.4 mg versus placebo under the intention-to-treat analysis. Race-specific estimates are displayed to the right, and sex-specific estimates are below the corresponding margins. External information is available only for marginal subgroups and not for the joint race-by-sex subgroups of interest. The ``Other'' race category in SURMOUNT-1 includes American Indian or Alaska Native, Native Hawaiian or Other Pacific Islander, and individuals reporting multiple races. Comparable results are not available from STEP~1 and STEP~2 using the same definition, and this category is therefore not shown. Abbreviation: CATE, conditional average treatment effect.}
    \label{fig:info}
\end{figure}

We illustrate our method by analyzing treatment effect heterogeneity in obesity pharmacotherapy. Assessing whether treatment effects differ across race and sex subgroups is important for evaluating the generalizability and equity of emerging anti-obesity medications, especially given the under-representation of males and racial minority groups in trials \citep{alsaqaaby2024sex}. We use SURMOUNT-1 \citep{jastreboff2022tirzepatide}, a phase~3 RCT of tirzepatide (Zepbound; Eli Lilly) for the treatment of obesity or overweight in adults without diabetes, to estimate CATEs for percentage weight loss across race-by-sex subgroups. Because sample sizes in some race-by-sex subgroups are sparse (Figure~\ref{fig:info}), detecting treatment heterogeneity using the internal trial alone can be difficult. We therefore leverage information from two previously conducted phase~3 RCTs of a related drug \citep{wilding2021once,davies2021semaglutide}, semaglutide (Wegovy; Novo Nordisk), STEP~1 (in adults without diabetes) and STEP~2 (in adults with diabetes). The available external information includes race-specific and sex-specific CATE estimates reported as point estimates and 95\% CIs (Figure~\ref{fig:info}). For example, among White participants, the estimated CATE for the 2.4 mg dose of semaglutide was a $-13.1\%$ change in body weight from baseline [95\% CI: $-14.1\%$, $-12.0\%$] in STEP~1 and $-7.22\%$ [95\% CI: $-8.61\%$, $-5.82\%$] in STEP~2. These differences align with well-documented differences in weight-loss drug response between diabetic and non-diabetic populations \citep{konwar2022efficacy}. This setting presents two key challenges: first, the external estimates are coarser than the race-by-sex subgroup effects of interest; second, these external estimates may be incompatible with their counterparts in SURMOUNT-1 due to differences in drugs and study populations, particularly diabetes status.

To our knowledge, existing methods do not integrate coarsened external causal effect estimates defined on marginal subgroups to improve estimation of more detailed subgroup-specific CATEs in an internal RCT. Our proposed JS estimator addresses this gap by allowing possible violations of compatibility between the internal and external trials with respect to causal effects at the marginal subgroup level. We show that, under certain conditions, the proposed estimator attains lower target quadratic risk than the internal-only estimator when estimating fine-grained subgroup CATEs, and that this result applies under general quadratic risk forms. We also provide a simple analytic variance estimator for the proposed JS estimator based on asymptotic theory that performs acceptably in practice and shows favorable performance relative to competing methods in simulation studies, along with a naive Wald-type confidence interval. 

The remainder of the paper is structured as follows. Section~2 reviews related work. Section~3 defines the problem, including notation and identification of causal effects. Section~4 presents the proposed method and outlines alternative methods for comparison. Section~5 evaluates performance through simulation studies and compares it with the alternative methods. Section~6 applies the proposed method to SURMOUNT-1 using coarsened external estimates from STEP~1 and STEP~2. Section~7 concludes with a discussion of key findings, limitations, and directions for future research.

\section{Related Work}

Integrating data from multiple sources is a well-established strategy in causal inference \citep{shi2023data}, particularly for improving the estimation of CATEs. Methods for leveraging information across RCTs vary with the type and granularity of external data \citep{brantner2023methods}. When only aggregate-level summaries of treatment effects, the most common form, are available, standard linear random-effects meta-analysis can be used. However, this approach assumes comparable analyses across studies, either through inclusion of the same treatment-covariate interaction terms \citep{kovalchik2013aggregate} or through reporting of subgroup-specific CATE estimates at the same level of granularity \citep{godolphin2023estimating}. Methods for leveraging external treatment effect estimates available only at a coarser level than the subgroup CATEs of interest remain underdeveloped. A recent contribution by \cite{van2023conditional} proposed a marginally constrained model that uses the ATE estimate from an external RCT as a constraint to assist CATE estimation in an internal observational study. While promising, this approach assumes that the external and internal studies arise from the same population and is limited to a single binary moderator.

Our proposed method builds on the literature of JS shrinkage estimators. Originally introduced by \cite{JamesStein1961}, the JS estimator achieves uniformly lower mean squared error (MSE) than the maximum likelihood estimator (MLE) when estimating multivariate normal means in dimensions greater than two. Since then,  this idea has been extended to a variety of settings. \cite{jennrich1986much} applied JS-type shrinkage to multiple linear regression. \cite{green1991james} later proposed combining an unbiased estimator with a potentially biased one, yielding an optimal linear combination of sample data and auxiliary information. More recently, \cite{rosenman2023combining} extended this framework to causal inference by developing a JS-type estimator that combines estimates from an RCT and an observational study without requiring parametric assumptions. A limitation of this method is that it assumes both studies estimate CATEs at the same level of granularity and that the underlying causal effects coincide. In a related prediction setting, \cite{han2024improving} introduced a JS estimator that shrinks an internal-only estimator toward one incorporating external summary information, often achieving lower prediction mean squared error (PMSE) across a range of heterogeneity levels between internal and external populations.

\section{Problem Formulation and Framework}

\subsection{Context and Notation of the Problem}

We consider two RCTs: an internal RCT with individual-level data (e.g., SURMOUNT-1) and an external RCT that provides only coarsened subgroup effect estimates (e.g., STEP~1 or STEP~2). In the internal study, we observe independent observations $(Y_i, \boldsymbol{X}_i, A_i)$ for $n$ participants, where $Y_i$ is a continuous outcome, $\boldsymbol{X}_i$ is a vector of $m$ categorical baseline pre-treatment covariates, and $A_i \in \{0,1\}$ denotes treatment assignment. Potential outcomes are denoted by $Y_i(1)$ and $Y_i(0)$ under treatment and control, respectively, following the potential outcome framework \citep{rubin1974estimating}. We assume that, in the internal population, the joint distribution of $(Y(1), Y(0), \boldsymbol{X}, A)$ is given by $F_I$. Let $\mathcal{G}$ denote the set of subgroups induced by the full cross-classification of the categorical covariates $\boldsymbol{X}$, and let $G = |\mathcal{G}|$ be the total number of such subgroups. For each $g \in \mathcal{G}$, defined by covariate profile $\boldsymbol{x}_g$, the target estimand is the fine-grained subgroup-specific CATE,
\begin{align*}
    \tau_I(g) = \mathbbm{E}_{F_I}\!\left[ Y(1) - Y(0) \mid \boldsymbol{X} = \boldsymbol{x}_g \right].
\end{align*}

For the external population, we assume an analogous joint distribution $F_E \big(Y(1), Y(0), \boldsymbol{X}, A\big)$. Let $\mathcal{G}_s$ denote the set of marginal subgroups for which external estimates are available, and let $q = |\mathcal{G}_s|$. For each subgroup $g_s \in \mathcal{G}_s$, the corresponding CATE is,
\begin{align*}
\gamma_E(g_s)
= \mathbbm{E}_{F_E}\!\left[ Y(1) - Y(0) \mid \boldsymbol{X}_{g_s} = \boldsymbol{x}_{g_s} \right],
\end{align*}
where $\boldsymbol{X}_{g_s} \subset \boldsymbol{X}$ denotes a proper subset of baseline covariates. We assume that the external study provides point estimates $\widehat{\gamma}_E(g_s)$ along with their estimated variances $\widehat{\mathrm{Var}}\!\left(\widehat{\gamma}_{E}(g_s)\right)$ from the external study. When the baseline covariate subset $\boldsymbol{X}_{g_s}$ is empty, $\gamma_E$ simplifies to the external ATE, that is, $\gamma_E = \mathbbm{E}_{F_E} [Y(1) - Y(0)]$. The analogous parameter in the internal population is denoted by $\gamma_I(g_s)$.

Let $\boldsymbol{\tau}_I = \left(\tau_I(g)\right)_{g \in \mathcal{G}} \in \mathbbm{R}^G$ denote the vector of internal fine-grained subgroup CATEs, and let $\widehat{\boldsymbol{\gamma}}_E
=
\big(\widehat{\gamma}_E(g_s)\big)_{g_s \in \mathcal{G}_s}
\in \mathbbm{R}^q$ denote the vector of available coarsened external subgroup CATE estimates from which information is borrowed. Our goal is to improve the efficiency of estimating $\boldsymbol{\tau}_I$ in the internal study by incorporating information from $\widehat{\boldsymbol{\gamma}}_E$. Specifically, we consider an arbitrary quadratic risk for an estimator $\widehat{\boldsymbol{\tau}}_I$, 
\begin{align}
\label{eq:risk}
R(\widehat{\boldsymbol{\tau}}_I) = \mathbbm{E}_{F_I}\!\bigl[(\widehat{\boldsymbol{\tau}}_I - \boldsymbol{\tau}_I)^{\prime} \boldsymbol{W}_n (\widehat{\boldsymbol{\tau}}_I - \boldsymbol{\tau}_I)\bigr],
\end{align}
which corresponds to a weighted sum of MSEs across subgroups. Here, $\boldsymbol{W}_n$ is a diagonal, positive semi-definite $G \times G$ weighting matrix. Section~4.3.1 provides examples of $\boldsymbol{W}_n$ and shows that the proposed JS estimator accommodates a broad class of quadratic risk criteria of the form \eqref{eq:risk}. A summary of notation is provided in Web Tables S1 and S2.

\subsection{Identification of causal effects}

For the internal study, we adopt standard causal assumptions typically satisfied in RCTs \citep{rubin1980randomization}: treatment exchangeability, $Y(a) \perp\!\!\!\perp A \mid \boldsymbol{X}$; positivity, $\mathbbm{P}(A = a \mid \boldsymbol{X}) > 0$ almost surely; and consistency, $Y = Y(a)$ when $A = a$. Under these assumptions, the fine-grained subgroup-specific CATE is identified by
\begin{align}
\mathbbm{E}_{F_I}[Y(1) - Y(0) \mid\boldsymbol{X} = \boldsymbol{x}_g] = \mathbbm{E}_{F_I}[Y \mid\boldsymbol{X} = \boldsymbol{x}_g, A = 1] - \mathbbm{E}_{F_I}[Y \mid\boldsymbol{X} = \boldsymbol{x}_g, A = 0],
\end{align}
with a derivation provided in Web Appendix~A.1.

For the external study, we assume $\widehat{\gamma}_E(g_s)$ is a consistent estimator of $\gamma_E(g_s)$ and that the external sample size $n_E$ is large relative to $n$. To incorporate these coarsened subgroup estimates, we introduce the following working compatibility condition:
\begin{condition}[Compatibility] \label{compatibility}
    $\boldsymbol{\gamma}_I = \boldsymbol{\gamma}_E$.
\end{condition}
Some estimators in Section~4 rely on Condition~\ref{compatibility}, but the proposed method does not. Since Condition~\ref{compatibility} concerns only marginal subgroup CATEs, it is weaker than standard data-integration assumptions such as strong ignorability of trial membership and mean exchangeability \citep{colnet2024causal}.

\section{Method}

\subsection{Unconstrained Estimator}

We begin by describing the OLS estimator, hereafter referred to as the unconstrained estimator, which uses only internal study data and does not incorporate external information.

\subsubsection{Method}

We fit a saturated linear regression model for the observed outcome that includes the treatment $A$, the full cross-classification of the categorical covariates $\boldsymbol{X}$, and all treatment-covariate interaction terms. Let $\boldsymbol{Y}$ denote the $n \times 1$ vector of observed outcomes, and let $\boldsymbol{X}^c$ denote the covariate $\boldsymbol{X}$ centered at its empirical mean $\bar{\boldsymbol{X}} = n^{-1}\sum_{i=1}^n \boldsymbol{X}_i$. Let $\boldsymbol{H}_n$ be the $n \times p$ design matrix formed by expanding $\boldsymbol{X}^c$ into all main effects and interactions, i.e., $\boldsymbol{H}_n = [\,\boldsymbol{h}_{X^c}(\boldsymbol{x}^c_1),\ldots,\boldsymbol{h}_{X^c}(\boldsymbol{x}^c_n)\,]^\prime$, where $\boldsymbol{h}_{X^c}(\cdot)$ denotes the corresponding expansion. Let $\boldsymbol{D}_A$ denote the $n\times n$ diagonal matrix with treatment assignments $\{a_1,\ldots,a_n\}$ on the diagonal. The full parameter vector $\boldsymbol{\theta}= (\boldsymbol{\alpha}^\prime,\boldsymbol{\beta}^\prime)^\prime$ comprises nuisance coefficients $\boldsymbol{\alpha}$, which capture outcome variation across covariate strata, and treatment heterogeneity coefficients $\boldsymbol{\beta}$, which capture how treatment effects vary across these strata, with both vectors of length $p$. The OLS estimator is
\begin{equation}
\label{eq:uc}
\widehat{\boldsymbol{\theta}}_{uc} = \underset{(\boldsymbol{\alpha},\,\boldsymbol{\beta})}{\arg\min}\;
\bigl(\boldsymbol{Y}-\boldsymbol{H}_n\boldsymbol{\alpha}-\boldsymbol{D}_A\boldsymbol{H}_n\boldsymbol{\beta}\bigr)^\prime
\bigl(\boldsymbol{Y}-\boldsymbol{H}_n\boldsymbol{\alpha}-\boldsymbol{D}_A\boldsymbol{H}_n\boldsymbol{\beta}\bigr).
\end{equation}
Let $\boldsymbol{L}_n$ denote the linear operator mapping the $p$ treatment heterogeneity coefficients to the $G$ fully stratified subgroup-specific CATEs. The resulting unconstrained CATE estimator is given by $\widehat{\boldsymbol{\tau}}_{uc} = \boldsymbol{L}_n\widehat{\boldsymbol{\beta}}_{uc}$. Consistency of OLS-based causal effect estimators under standard regularity conditions is well known \citep{yang2001efficiency,lin2013agnostic}; however, finite-sample unbiasedness is not guaranteed. We establish unbiasedness in this setting.

\begin{theorem}
\label{thm:unbiased}
For each subgroup $g\in\mathcal{G}$, if the population distributions of the potential outcomes have finite means and variances, then $\widehat{\boldsymbol{\tau}}_{uc}$ is an unbiased estimator of $\boldsymbol{\tau}_{I}$.
\end{theorem}

The proof is provided in Web Appendix~A.2, where we extend the argument of \citet{schochet2010regression} from the unadjusted regression for ATE estimation to the saturated regression considered here for subgroup-specific CATE estimation. 

By residualizing both the outcome $\boldsymbol{Y}$ and the treatment-covariate interaction regressors $\boldsymbol{D}_A\boldsymbol{H}_n$ with respect to the nuisance regressors $\boldsymbol{H}_n$, the unconstrained estimator of the treatment heterogeneity coefficients admits the closed-form expression
\(
\widehat{\boldsymbol{\beta}}_{uc}
= (\boldsymbol{K}_n^{\prime}\boldsymbol{K}_n)^{-1}\boldsymbol{K}_n^{\prime}\boldsymbol{Y},
\)
where $\boldsymbol{K}_n = \boldsymbol{M}_n\boldsymbol{D}_A\boldsymbol{H}_n$ is an $n \times p$ matrix and $\boldsymbol{M}_n = \boldsymbol{I}_n -\boldsymbol{H}_n(\boldsymbol{H}_n^{\prime}\boldsymbol{H}_n)^{-1}\boldsymbol{H}_n^{\prime}$ is the $n \times n$ projection matrix onto the orthogonal complement of the column space of the nuisance regressors $\boldsymbol{H}_n$, with $\boldsymbol{I}_n$ denoting the $n\times n$ identity matrix.

\begin{example}
\label{ex:two-binary}
In Section~4, we consider a running example in which the target estimand is the CATEs defined by the joint values of two binary moderators, $X_1$ and $X_2$, yielding four subgroups. We center the moderators as $X^c_{1,i} = X_{1,i}-\bar{X}_1$ and $X^c_{2,i} = X_{2,i}-\bar{X}_2$ for $i=1,\dots,n$. Here, $\bar{X}_1 = n^{-1}\sum_{i=1}^n X_{1,i}$ and $\bar{X}_2 = n^{-1}\sum_{i=1}^n X_{2,i}$ denote the empirical means of the moderators $X_1$ and $X_2$, respectively. The mapping $\boldsymbol{h}_{X^c}(X^c_{1,i},X^c_{2,i}) = (1,\,X^c_{1,i},\,X^c_{2,i},\,X^c_{1,i}X^c_{2,i})^\prime$ expands the $m=2$ centered covariates into the full set of $p=4$ regressors. Stacking these row vectors yields the $n\times 4$ design matrix $\boldsymbol{H}_n$. The unconstrained estimator then takes the form given in Equation \eqref{eq:uc}, with both $\boldsymbol{\alpha}$ and $\boldsymbol{\beta}$ of dimension $p=4$. For the four fully stratified subgroups, the mapping from the treatment heterogeneity
coefficients $\widehat{\boldsymbol{\beta}}_{uc}$ to the subgroup CATE
estimates $\widehat{\boldsymbol{\tau}}_{uc}$ is $\widehat{\boldsymbol{\tau}}_{uc} = \boldsymbol{L}_n\,\widehat{\boldsymbol{\beta}}_{uc}$, where
\[
\boldsymbol{L}_n=
\begin{bmatrix}
1 & -\bar X_1 & -\bar X_2 & (-\bar X_1)(-\bar X_2)\\
1 & -\bar X_1 & 1-\bar X_2 & (-\bar X_1)(1-\bar X_2)\\
1 & 1-\bar X_1 & -\bar X_2 & (1-\bar X_1)(-\bar X_2)\\
1 & 1-\bar X_1 & 1-\bar X_2 & (1-\bar X_1)(1-\bar X_2)
\end{bmatrix}.
\]
\end{example}

\subsubsection{Asymptotic Distribution} This section describes the asymptotic behavior of the unconstrained estimator of the treatment heterogeneity coefficients $\widehat{\boldsymbol{\beta}}_{uc}$ as $n\to\infty$. The corresponding results for the CATE estimator $\widehat{\boldsymbol{\tau}}_{uc}$ follow directly via the linear transformation $\boldsymbol{L}_n$. We write the fully stratified outcome model as $\boldsymbol{Y} = \boldsymbol{H}_n \boldsymbol{\alpha} + \boldsymbol{D}_A \boldsymbol{H}_n \boldsymbol{\beta} + \boldsymbol{\varepsilon}$, which defines the regression error $\boldsymbol{\varepsilon}$; as shown in Web Appendix~A.2, $\mathbbm{E}_{F_I}[\boldsymbol{\varepsilon}\mid \boldsymbol{X}, A] = \boldsymbol{0}$. Under standard regularity conditions,
\begin{equation}
\label{eq:uca}
\sqrt{n}\,(\widehat{\boldsymbol{\beta}}_{uc}-\boldsymbol{\beta})
\;\overset{d}{\to}\;
\mathcal{Z}_{uc},
\end{equation}
where $\mathcal{Z}_{uc} \sim \mathcal{N}(\boldsymbol{0}, \boldsymbol{\Sigma}_{uc})$, with $\boldsymbol{\Sigma}_{uc}
= \lim_{n \to \infty}
n (\boldsymbol{K}_n^{\prime}\boldsymbol{K}_n)^{-1}
\boldsymbol{K}_n^{\prime}\boldsymbol{\Omega}\boldsymbol{K}_n
(\boldsymbol{K}_n^{\prime}\boldsymbol{K}_n)^{-1}.$ Here, $\boldsymbol{\Omega}=\operatorname{Var}_{F_I}(\boldsymbol{\varepsilon}\mid\boldsymbol{X},A)$ is diagonal under independence across units, with unequal diagonal elements $\operatorname{Var}_{F_I}(\varepsilon_i\mid\boldsymbol{X}_i,A_i)$, allowing for heteroskedasticity. We estimate the finite-sample variance-covariance matrix $\widehat{\operatorname{Var}}\!\left(\widehat{\boldsymbol{\beta}}_{uc}\right)$ via a Huber-White sandwich estimator \citep{white1980heteroskedasticity}. Following \citet{samii2012equivalencies}, we use the heteroskedasticity-consistent version 2 (HC2) adjustment, which coincides with Neyman's finite-population variance for the ATE under unadjusted regression. Specifically,
\[
\widehat{\operatorname{Var}}\!\left(\widehat{\boldsymbol{\beta}}_{uc}\right)
=
(\boldsymbol{K}_n^{\prime}\boldsymbol{K}_n)^{-1}\boldsymbol{K}_n^{\prime}
\widehat{\boldsymbol{\Omega}}
\boldsymbol{K}_n
(\boldsymbol{K}_n^{\prime}\boldsymbol{K}_n)^{-1},
\]
where $\widehat{\boldsymbol{\Omega}}$ is a $n \times n$ diagonal matrix with $i$th diagonal element $\widehat{\varepsilon}_i^{\,2}/(1-h_{ii})$, $\widehat{\varepsilon}_i$ is the $i$th regression residual, and $h_{ii}$ is the $i$th diagonal element of the hat matrix associated with the joint design matrix $[\,\boldsymbol{H}_n \;\; \boldsymbol{D}_A\boldsymbol{H}_n\,]$. The asymptotic variance–covariance matrix $\boldsymbol{\Sigma}_{uc}$ is estimated by the plug-in estimator $\widehat{\boldsymbol{\Sigma}}_{uc}
=
n\,\widehat{\operatorname{Var}}(\widehat{\boldsymbol{\beta}}_{uc})$.

\subsection{Constrained Estimator}

This section introduces the constrained least squares (CLS) estimator, hereafter, the constrained estimator, which incorporates internal data while restricting the parameter space to align with the coarsened external subgroup CATE estimates $\widehat{\boldsymbol{\gamma}}_E$.

\subsubsection{Method}
Let $\widehat{\boldsymbol{\theta}}_{c} = (\widehat{\boldsymbol{\alpha}}_{c}^{\prime},\widehat{\boldsymbol{\beta}}_{c}^{\prime})^{\prime}$ denote the constrained estimator of the regression coefficients, obtained by minimizing the objective function in Equation~\eqref{eq:uc} subject to linear constraints,
\begin{equation*}
\widehat{\boldsymbol{\theta}}_{c}
= \underset{(\boldsymbol{\alpha},\,\boldsymbol{\beta})}{\arg\min}\;
\bigl(\boldsymbol{Y} - \boldsymbol{H}_n\boldsymbol{\alpha} - \boldsymbol{D}_A\boldsymbol{H}_n\boldsymbol{\beta}\bigr)^{\prime}
\bigl(\boldsymbol{Y} - \boldsymbol{H}_n\boldsymbol{\alpha} - \boldsymbol{D}_A\boldsymbol{H}_n\boldsymbol{\beta}\bigr)
\quad \text{subject to} \quad
\boldsymbol{C}_n\boldsymbol{\beta} = \widehat{\boldsymbol{\gamma}}_{E},
\end{equation*}
where $\boldsymbol{C}_n$ is a full-rank $q\times p$ matrix ($q<p$) that imposes $q$ linearly independent restrictions on the $p$ treatment heterogeneity coefficients $\boldsymbol{\beta}$. The number of restrictions is smaller than the number of heterogeneity coefficients because the external estimates correspond to marginal subgroups, whereas the internal model quantifies CATEs for fine-grained subgroups defined by additional interactions among baseline covariates $\boldsymbol{X}$. As in the unconstrained one, the CATE estimator is obtained by applying the same linear transformation $\boldsymbol{L}_n$, $\widehat{\boldsymbol{\tau}}_{c} = \boldsymbol{L}_n\,\widehat{\boldsymbol{\beta}}_{c}.$

A closed-form expression for $\widehat{\boldsymbol{\beta}}_{c}$ follows from the method of Lagrange multipliers:
\begin{align}
\label{eq:ce}
\widehat{\boldsymbol{\beta}}_{c} = \widehat{\boldsymbol{\beta}}_{uc} + \boldsymbol{B}_{n}\bigl(\widehat{\boldsymbol{\gamma}}_{E}
      -\boldsymbol{C}_{n}\widehat{\boldsymbol{\beta}}_{uc}\bigr),
\end{align}
where $\boldsymbol{B}_{n}$ is a $p \times q$ matrix that maps the $q$-dimensional space marginal external estimates to the $p$-dimensional space treatment heterogeneity parameters. Specifically, $\boldsymbol{B}_{n}=(\boldsymbol{K}_{n}^{\prime}\boldsymbol{K}_{n})^{-1}
\boldsymbol{C}_n^{\prime}
\bigl[\boldsymbol{C}_n(\boldsymbol{K}_{n}^{\prime}\boldsymbol{K}_{n})^{-1}\boldsymbol{C}_n^{\prime}\bigr]^{-1}$.

\begin{remark}
    The constrained estimator follows the idea of \citet{han2024improving}, but the restrictions here are applied only to the treatment heterogeneity parameters $\boldsymbol{\beta}$, rather than to the full regression parameter vector $\boldsymbol{\theta}$. We note that other approaches could also be used, such as empirical likelihood (EL) methods \citep{qin1994empirical}.
\end{remark}

\addtocounter{example}{-1}
\begin{example}[continued]
Consider the running example where external estimates are available for both the $X_1$-defined and $X_2$-defined subgroup CATEs. Collect these estimates in a $q \times 1$ vector
\(
\widehat{\boldsymbol{\gamma}}_E 
= \big(\widehat{\gamma}_E(X_1=0),\; \widehat{\gamma}_E(X_1=1),\; 
       \widehat{\gamma}_E(X_2=0)\big)^{\prime},
\)
with $q=3$. The corresponding linear mapping can be written as $\boldsymbol{C}_n\boldsymbol{\beta} 
= \widehat{\boldsymbol{\gamma}}_E$, where $\boldsymbol{C}_n$ is a full-rank $q \times p$ matrix with $p=4$. The number of restrictions $q$ is smaller than the length of the treatment heterogeneity coefficient vector $p$ because the internal model includes an additional interaction between $X_1$ and $X_2$ that defines finer-grained subgroups. Specifically,
\[
\boldsymbol{C}_n =
\begin{bmatrix}
    1 & -\bar{X}_1 & 0 & 0\\
    1 & 1-\bar{X}_1 & 0 & 0 \\
    1 & 0 & -\bar{X}_2 & 0
\end{bmatrix}.
\]
Empirical centering of $\boldsymbol{X}$ yields a simpler form of $\boldsymbol{C}_n$.
\end{example}

\subsubsection{Asymptotic Distribution} This section describes the asymptotic behavior of the constrained estimator $\widehat{\boldsymbol{\beta}}_{c}$ as $n\to\infty$. The corresponding results for the CATE estimator $\widehat{\boldsymbol{\tau}}_{c}$ follow directly from the linear transformation $\boldsymbol{L}_n$. Given the relationship between the constrained and unconstrained estimators in Equation~\eqref{eq:ce}, we have $\widehat{\boldsymbol{\beta}}_{c}
\xrightarrow{p}
\boldsymbol{\beta}_{c}
=
\boldsymbol{\beta}
+
\boldsymbol{B}\bigl(\boldsymbol{\gamma}_{E}
-
\boldsymbol{C}\boldsymbol{\beta}\bigr),$
where $\boldsymbol{C} = \lim_{n \to \infty} \boldsymbol{C}_n$ and $\boldsymbol{B} = \lim_{n \to \infty} \boldsymbol{B}_n$. Although $\boldsymbol{C}_n$ and $\boldsymbol{B}_n$ are data-dependent, their dimensions are fixed. Under the compatibility working condition~\ref{compatibility}, $\boldsymbol{\beta}_{c}=\boldsymbol{\beta}$, so the constrained estimator is consistent. By the relationship in Equation~\eqref{eq:ce}, under standard regularity conditions,
\begin{align}
\label{eq:ca}
\sqrt{n}\,(\widehat{\boldsymbol{\beta}}_{c}-\boldsymbol{\beta}_{c})
\;\overset{d}{\to}\; \mathcal{Z}_{c},
\end{align}
where $\mathcal{Z}_{c} = (\boldsymbol{I}_{p}-\boldsymbol{B}\boldsymbol{C})\,\mathcal{Z}_{uc}$, with $\mathcal{Z}_{uc}$ defined in Equation~\eqref{eq:uca}. Consequently,
\[\mathcal{Z}_{c} \sim \mathcal{N}\!\left(\boldsymbol{0},
(\boldsymbol{I}_{p}-\boldsymbol{B}\boldsymbol{C})\boldsymbol{\Sigma}_{uc}
(\boldsymbol{I}_{p}-\boldsymbol{B}\boldsymbol{C})^{\prime}\right).\] See Section~4.1.2 for the estimation of $\boldsymbol{\Sigma}_{uc}$. The asymptotic variance of $\widehat{\boldsymbol{\beta}}_{c}$ is independent of the true coarsened external CATEs $\boldsymbol{\gamma}_E$ and is no larger than that of the unconstrained estimator. Under the compatibility condition~\ref{compatibility}, this reduction implies an efficiency gain.

\subsection{Proposed James-Stein Estimator}

The unconstrained estimator is consistent and unbiased. When the compatibility condition~\ref{compatibility} holds, the constrained estimator is also consistent and attains lower asymptotic variance. When the condition is violated, however, the constrained estimator may incur asymptotic bias. These properties highlight a bias–variance tradeoff: while the constrained estimator can improve efficiency when the external coarsened CATEs are compatible with their internal counterparts, it may perform poorly otherwise. Motivated by this tradeoff, we seek an estimator that remains consistent while shrinking towards the constrained estimator. This section introduces our proposed James-Stein (JS) shrinkage estimator.

\subsubsection{Method} 
We begin by defining the original JS estimator (without the positive-part adjustment) for the treatment heterogeneity coefficients, $\widehat{\boldsymbol{\beta}}_{js}$, as a scalar-weighted average of the unconstrained estimator $\widehat{\boldsymbol{\beta}}_{uc}$ and the constrained estimator $\widehat{\boldsymbol{\beta}}_{c}$:
\begin{align}
\label{eq:js}
\widehat{\boldsymbol{\beta}}_{js} &= \widehat{\omega} \widehat{\boldsymbol{\beta}}_{uc} + (1 - \widehat{\omega}) \widehat{\boldsymbol{\beta}}_{c}.
\end{align}
The data-adaptive shrinkage factor $\widehat{\omega}$ is defined as
\begin{align}
\label{eq:w}
\widehat{\omega}
=
1
-
\frac{\nu_n}{
n\,
(\widehat{\boldsymbol{\tau}}_{uc}-\widehat{\boldsymbol{\tau}}_{c})^{\prime}
\boldsymbol{W}_n
(\widehat{\boldsymbol{\tau}}_{uc}-\widehat{\boldsymbol{\tau}}_{c})
},   
\end{align}
where $\nu_n$ is a data-dependent tuning parameter. The quadratic form in the denominator is analogous to the target quadratic risk in Equation \eqref{eq:risk}; specifically, for any CATE estimator $\widehat{\boldsymbol{\tau}}_I$,
$
R(\widehat{\boldsymbol{\tau}}_I)
=
\mathbbm{E}_{F_I}\!\left[
(\widehat{\boldsymbol{\tau}}_I-\boldsymbol{\tau}_I)^{\prime}
\boldsymbol{W}_n
(\widehat{\boldsymbol{\tau}}_I-\boldsymbol{\tau}_I)
\right].$ Thus, the choice of $\boldsymbol{W}_n$ determines the form of shrinkage. In the main text, we restrict attention to prevalence weighting, which uses a diagonal matrix with entries given by the estimated subgroup prevalence, assigning greater weight to larger subgroups. See Web Appendix B for more examples. Using the same linear transformation $\boldsymbol{L}_n$ that maps the $p$ treatment heterogeneity coefficients to the $G$ fine-grained subgroup CATEs, as defined earlier in Section~4.1, the corresponding fine-grained subgroup CATE estimator is then given by $\widehat{\boldsymbol{\tau}}_{js}
=
\boldsymbol{L}_n\widehat{\boldsymbol{\beta}}_{js}$.

The following theorem compares the target quadratic risk of the original JS estimator $\widehat{\boldsymbol{\tau}}_{js}$ with that of the unconstrained estimator $\widehat{\boldsymbol{\tau}}_{uc}$. To establish this result, we additionally assume normality of the regression errors. This assumption is imposed only to derive finite-sample risk dominance via Stein’s identity \citep{stein1981estimation}. Similar normality assumptions have been used in prior analyses of JS–type estimators \citep{jennrich1986much,green1991james,han2024improving}.

\begin{theorem}
\label{thm:js}
Assume that the regression errors satisfy $\boldsymbol{\varepsilon} \sim N(0,\, \boldsymbol{\Omega})$, so that $\widehat{\boldsymbol{\beta}}_{uc} \sim N(\boldsymbol{\beta},\, \boldsymbol{\Sigma}_{uc}/n)$ in finite samples. Define the $G \times G$ matrix $\boldsymbol{P} = \mathbbm{E}_{F_I}\!\left(\boldsymbol{W}_n\boldsymbol{L}_n\boldsymbol{\Sigma}_{uc}\boldsymbol{C}_n^{\prime}\boldsymbol{B}_n^{\prime}\boldsymbol{L}_n^{\prime}\right)$, and suppose that $\mathrm{tr}(\boldsymbol{P})/\|\boldsymbol{P}\| > 2$ and $0 < \nu_n < 2\{\mathrm{tr}(\boldsymbol{P}) - 2\|\boldsymbol{P}\|\}$. Then, for all $\boldsymbol{\tau}_{I}$, the JS estimator satisfies $R(\widehat{\boldsymbol{\tau}}_{js}) \le R(\widehat{\boldsymbol{\tau}}_{uc})$. Moreover, the target risk of $\widehat{\boldsymbol{\tau}}_{js}$ is minimized at $\nu_n = \mathrm{tr}(\boldsymbol{P}) - 2\|\boldsymbol{P}\|$, where $\|\cdot\|$ denotes the matrix spectral norm.
\end{theorem}

The proof of Theorem~\ref{thm:js} extends \citet{han2024improving} to the causal inference setting considered here. Specifically, the target risk shifts from the PMSE associated with the full regression parameter to a general quadratic loss involving only the treatment heterogeneity coefficients, and homoscedasticity is not required. Details are given in Web Appendix~A.3. In practice, a data-driven choice may be obtained by replacing the theoretical matrix $\boldsymbol{P}$ with its plug-in estimator $\widehat{{\boldsymbol{P}}} = \boldsymbol{W}_n \boldsymbol{L}_n\widehat{\boldsymbol{\Sigma}}_{uc}\boldsymbol{C}_n^{\prime}\boldsymbol{B}_n^{\prime}\boldsymbol{L}_n^{\prime}$ and setting the tuning parameter to $\nu_n = \mathrm{tr}(\widehat{{\boldsymbol{P}}}) - 2\|\widehat{{\boldsymbol{P}}}\|$.

\begin{remark}
\label{remark:sufficient}
    A sufficient condition to ensure that $\mathrm{tr}(\boldsymbol{P})/\|\boldsymbol{P}\| > 2$ is that $\mathrm{rank}(\boldsymbol{P}) > 2.$ Equivalently, both the rank of $\boldsymbol{C}_n$ (i.e., $q$) and the rank of $\boldsymbol{L}_n^{\prime}\boldsymbol{W}_n\boldsymbol{L}_n$ must be at least three. This implies that at least three linearly independent restrictions on $\boldsymbol{\beta}$ are required (i.e., the ATE alone is insufficient), and that the target risk must involve at least three fine-grained subgroup CATEs. A related observation is made in \citet{hansen2016efficient}.
\end{remark}

The original JS estimator in Equation~\eqref{eq:js} suffers from a singularity when $\widehat{\boldsymbol{\tau}}_{uc}-\widehat{\boldsymbol{\tau}}_{c}=\boldsymbol{0}$. To address this issue, \cite{baranchik1970family} proposed a positive-part adjustment on the shrinkage factor. The resulting positive-part JS estimator is defined as
\begin{align}
\label{eq:js+}
\widehat{\boldsymbol{\beta}}_{js^+} = \widehat{\omega}^+ \widehat{\boldsymbol{\beta}}_{uc} + (1 - \widehat{\omega}^+) \widehat{\boldsymbol{\beta}}_{c},
\end{align}
where $\widehat{\omega}^+ = \max(0, \widehat{\omega})$. Throughout, we adopt the positive-part version, denoted as $\widehat{\boldsymbol{\beta}}_{js^+}$, along with its corresponding CATE estimator $\widehat{\boldsymbol{\tau}}_{js^+}$.

\begin{remark}
    The positive-part adjustment yields a more stable and interpretable estimator. When the discrepancy between $\widehat{\boldsymbol{\tau}}_{uc}$ and $\widehat{\boldsymbol{\tau}}_{c}$ is large, the shrinkage factor $\widehat{\omega}^+ $ approaches $1$, assigning more weight on the unconstrained estimator $\widehat{\boldsymbol{\beta}}_{uc}$. In contrast, when the data strongly support the compatibility condition \ref{compatibility} such that $\widehat{\boldsymbol{\tau}}_{uc}-\widehat{\boldsymbol{\tau}}_{c} \to  \boldsymbol{0}$, the estimator relies primarily on the constrained estimator $\widehat{\boldsymbol{\beta}}_{c}$. 
\end{remark}

\subsubsection{Asymptotic Distribution}

For our proposed JS-type estimator, a key unresolved challenge is positive-part, data-adaptive weighting schemes, which lead to non-regular asymptotics \citep{casella2012shrinkage}. This non-regularity precludes the use of standard resampling methods, such as the bootstrap, that are commonly used in the JS literature \citep{yi1991estimating, rosenman2023combining}. Our goal is to develop a practically reliable variance estimator for our proposed JS estimator that is grounded in asymptotic theory and admits a closed-form expression. To this end, we follow the framework of \citet{chen2009shrinkage}, adapting it to our setting; their CATE shrinkage estimator is included for comparison in Section~4.4. We study the asymptotic behavior of the proposed JS estimator for the treatment heterogeneity coefficients, $\widehat{\boldsymbol{\beta}}_{js^+}$, as $n \to \infty$. The corresponding asymptotic results for the subgroup CATE estimator follow directly from the linear transformation $\boldsymbol{L}_n$. The normality assumption imposed for risk comparison in Theorem~\ref{thm:js} is not required for the analysis in this section.

When the compatibility condition~\ref{compatibility} holds, the limiting distribution of $\widehat{\boldsymbol{\beta}}_{js^+}$ is non-normal; see Web Appendix~C for details. When the compatibility condition \ref{compatibility} fails, so that $\boldsymbol{\gamma}_I\neq\boldsymbol{\gamma}_E$ and $\boldsymbol{\beta}_{c}\neq\boldsymbol{\beta}$, the shrinkage weight factor $\widehat{\omega} \xrightarrow{p} 1$, and the same holds for $\widehat{\omega}^+$. The reason is that, under incompatibility, the quadratic form in the denominator converges to a positive constant, and the scaling by $n$ drives the shrinkage factor to one (Equation~\eqref{eq:w}). As a consequence, the proposed estimator is asymptotically equivalent to the unconstrained estimator. Following \cite{chen2009shrinkage}, we therefore seek a small-sample–adjusted variance estimator rather than directly using the variance estimator of the unconstrained estimator. Define $\widetilde{\boldsymbol{\beta}}_{js}^{(n)}$ as a perturbation of the true parameter $\boldsymbol{\beta}$ by an $o_p(1)$ term involving population quantities: $\widetilde{\boldsymbol{\beta}}_{js}^{(n)}
=
\boldsymbol{\beta}
+
\{\nu/[n(\boldsymbol{\Delta}^{\prime}\boldsymbol{\Gamma}\boldsymbol{\Delta})]\}
\boldsymbol{\Delta}$, where $\nu_n\xrightarrow{p}\nu$, $\widehat{\boldsymbol{\Delta}}\xrightarrow{p}\boldsymbol{\Delta}$, and $\boldsymbol{L}_n^{\prime}\boldsymbol{W}_n\boldsymbol{L}_n\xrightarrow{p}\boldsymbol{\Gamma}$ defining the probability limits of the corresponding sample quantities. Here, $\boldsymbol{\Gamma}$ is a $p\times p$ matrix. By construction, $\widetilde{\boldsymbol{\beta}}_{js}^{(n)} \xrightarrow{p}
\boldsymbol{\beta}$. A first-order Taylor expansion of the original JS estimator $\widehat{\boldsymbol{\beta}}_{js}$ around $\widetilde{\boldsymbol{\beta}}_{js}^{(n)}$ yields
\begin{align}
\label{eq:jsain}
    \sqrt{n}\bigl(\widehat{\boldsymbol{\beta}}_{js}
    - \widetilde{\boldsymbol{\beta}}_{js}^{(n)}\bigr)
    = \mathcal{G}_{js}\,
    \sqrt{n}\bigl(\widehat{\boldsymbol{\beta}}_{uc} - \boldsymbol{\beta}\bigr)
    + o_p(1),
\end{align}
where $\mathcal{G}_{js}
= \boldsymbol{I}_p
- \{\nu/[n(\boldsymbol{\Delta}^\prime\boldsymbol{\Gamma}\boldsymbol{\Delta})]\}\,\boldsymbol{B}\boldsymbol{C}
+ \{2\nu/[n(\boldsymbol{\Delta}^\prime\boldsymbol{\Gamma}\boldsymbol{\Delta})^{2}]\}\,
\boldsymbol{\Delta}\boldsymbol{\Delta}^\prime\boldsymbol{\Gamma}\boldsymbol{B}\boldsymbol{C}$. The latter two terms vanish as $n \to \infty$, implying $\mathcal{G}_{js}$ converges to $\boldsymbol{I}_p$. The same conclusion applies to the positive–part estimator $\widehat{\boldsymbol{\beta}}_{js^+}$. 

In practice, the compatibility condition~\ref{compatibility} is not identifiable from the observed data. Following \citet{chen2009shrinkage}, one could use the variance estimator derived under incompatibility, $n^{-1}\widehat{\mathcal{G}}_{js}\widehat{\boldsymbol{\Sigma}}_{uc}\widehat{\mathcal{G}}_{js}^{\prime}$, uniformly, where $\widehat{\mathcal{G}}_{js}$ is the plug-in estimator of $\mathcal{G}_{js}$ in Equation~\eqref{eq:jsain}. However, in our setting this approach can be numerically unstable when $\widehat{\boldsymbol{\tau}}_{uc}-\widehat{\boldsymbol{\tau}}_{c}$ is extremely close to zero, as the denominator of the shrinkage factor $\widehat{\omega}$ in Equation~\eqref{eq:w}, leading to an ill-conditioned variance estimate. To address this issue, we adopt a data-adaptive, piecewise variance estimator: when $\widehat{\omega}^+=0$, we use the variance estimator of the constrained estimator; otherwise, we use $n^{-1}\widehat{\mathcal{G}}_{js}\widehat{\boldsymbol{\Sigma}}_{uc}\widehat{\mathcal{G}}_{js}^{\prime}$. The condition $\widehat{\omega}^+=0$ is equivalent to $n(\widehat{\boldsymbol{\tau}}_{uc}-\widehat{\boldsymbol{\tau}}_{c})^{\prime}\boldsymbol{W}_n(\widehat{\boldsymbol{\tau}}_{uc}-\widehat{\boldsymbol{\tau}}_{c})\le\nu_n$. Estimation of $\boldsymbol{\Sigma}_{uc}$ is described in Section~4.1.2. Similar data-adaptive variance constructions, in which the variance estimator is defined piecewise to ensure numerical stability in near-singular settings, have been adopted in the shrinkage literature \citep{boss2025mediation}.

The construction of CIs for JS-type estimators remains an open problem \citep{casella2012shrinkage}. Many existing methods target empirical Bayes (EB) coverage by accounting for repeated sampling of both the observed data and the underlying true effects, which is strictly weaker than standard frequentist coverage \citep{armstrong2022robust}. Procedures that achieve reliable frequentist coverage for JS-type estimators are comparatively scarce \citep{rosenman2023combining}. We therefore report naive Wald-type confidence intervals based on the proposed variance estimator, without bias adjustment.

\subsection{Other Forms of Shrinkage Estimators}

We consider two alternative shrinkage estimators that adaptively incorporate external coarsened CATE estimates, for comparison with our proposed JS estimator. In contrast to our method, their forms are invariant to the choice of the weighting matrix $\boldsymbol{W}_n$ defining the target risk for the subgroup CATEs. The first is an adaptive EB estimator \citep{mukherjee2008exploiting}, which takes the form of a matrix-weighted average of the unconstrained estimator $\widehat{\boldsymbol{\beta}}_{uc}$ and the constrained estimator $\widehat{\boldsymbol{\beta}}_{c}$, in contrast to the scalar-weighted average used in Equation~\eqref{eq:js+}. The second is a generalized ridge estimator that uses the variance estimate $\widehat{\mathrm{Var}}(\widehat{\boldsymbol{\gamma}}_E)$ to govern the strength of penalization of the point estimate $\widehat{\boldsymbol{\gamma}}_E$. Full details for both estimators are provided in Web Appendix~D.

\section{Simulation Study}

We conduct a simulation study with two main objectives: (1) to compare the target risk for the fine-grained subgroup CATE vector $\boldsymbol{\tau}_I$ across methods and to provide simulation-based validation of Theorem~\ref{thm:js} for the proposed JS estimator; and (2) to assess the performance of the proposed analytic variance estimator for the JS estimator.

\subsection{Simulation Settings}

We mirror the running example in Section~4 (Example~1), where subgroup CATEs are jointly defined by two binary baseline covariates, $X_1$ and $X_2$. The design is as follows:

\begin{enumerate}
    \item \textbf{External inputs.} We assume access to coarsened external information in the form of marginal subgroup CATE estimates defined separately by $X_1$ and by $X_2$, $\widehat{\boldsymbol{\gamma}}_E 
    = \big(\widehat{\gamma}_E(X_1=0),\; \widehat{\gamma}_E(X_1=1),\; 
           \widehat{\gamma}_E(X_2=0)\big)^{\prime}$, together with their corresponding variance estimates.
    
    \item \textbf{Internal population model.} We set the internal sample size to $n=500$ and generate $X_1$ and $X_2$ independently from Bernoulli distributions. Potential outcomes are
    \[
    Y(0) = \eta_0 + \eta_1 X_1 + \eta_2 X_2 + \eta_{12} X_1 X_2 + \varepsilon_{(0)}, 
    \quad \varepsilon_{(0)} \sim \mathcal{N}(0,\sigma^2_{(0)}),
    \]
    \[
    Y(1) = \zeta_0 + \zeta_1 X_1 + \zeta_2 X_2 + \zeta_{12} X_1 X_2 + \varepsilon_{(1)}, 
    \quad \varepsilon_{(1)} \sim \mathcal{N}(0,\sigma^2_{(1)}).
    \]
    Parameter values are given in Web Appendix~E. The observed outcome is constructed as $Y = (1 - A)Y(0) + A Y(1)$, where treatment is assigned as $A \sim \operatorname{Bernoulli}(0.5)$. This specification allows for heteroskedasticity in the regression error of the observed outcome.

    \item \textbf{Compatibility.} To evaluate sensitivity to violations of the compatibility condition~\ref{compatibility}, we define a incompatibility index for a marginal subgroup $g_s$ as $e(g_s) = (\gamma_E(g_s) - \gamma_I(g_s)) / \gamma_I(g_s)$, which represents the relative difference between the true marginal subgroup CATE in the external and internal populations. We impose a common incompatibility index $e$ across subgroups defined by $X_1$ and $X_2$, varying $e$ from $0$ to $0.1$ in increments of $0.002$. For each value of $e$, we generate $5{,}000$ Monte Carlo replications, indexed by $r=1,\ldots,5000$.
\end{enumerate}

The target risk for the fine-grained subgroup CATE vector jointly defined by $X_1$ and $X_2$  is
$
R(\widehat{\boldsymbol{\tau}}_I)
=
\frac{1}{5000}\sum_{r=1}^{5000}\sum_{g=1}^{4} \widehat{w}_{r,g}\,\bigl(\widehat{\tau}_I(r,g)-\tau_I(g)\bigr)^2,$
where $\widehat{w}_{r,g}$ denotes the $(g,g)$ diagonal entry of the weighting matrix $\boldsymbol{W}_n$ in the target quadratic risk (see Equation~\eqref{eq:risk}) for replication $r$. In the main text, we use prevalence weighting, with $\hat{w}_{r,g}$ equal to the empirical prevalence of subgroup $g$ in replication $r$. For each fine-grained subgroup-specific CATE $\tau_I(g)$, we report the empirical coverage of the naive Wald-type CIs.

\subsection{Simulation Results}

\begin{figure}
    \centering
    \includegraphics[width=1\linewidth]{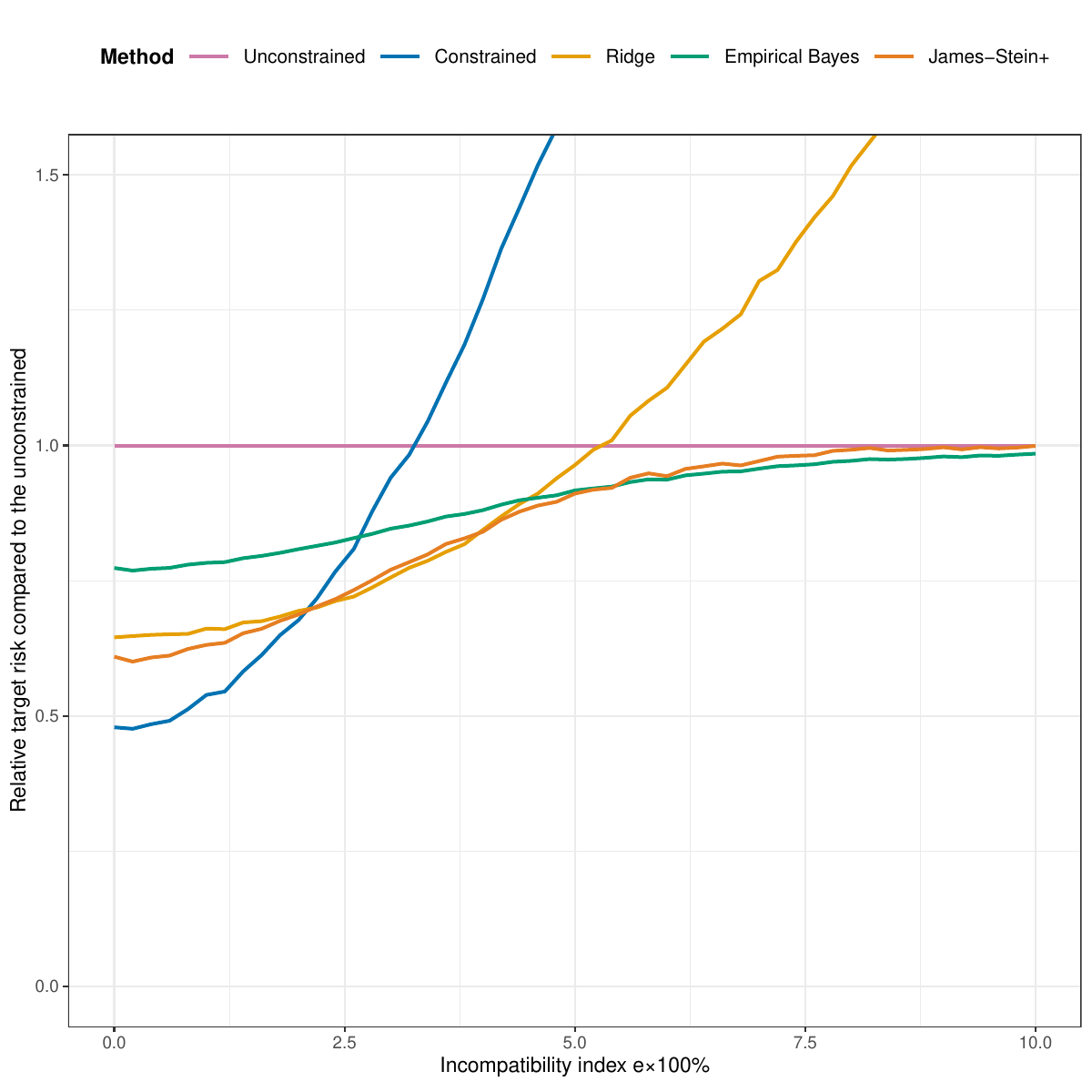}
    \caption{\textbf{Target risk comparison under prevalence weighting for the fine-grained subgroup-specific CATEs vector jointly defined by $X_1$ and $X_2$, expressed relative to the unconstrained estimator.} The horizontal axis denotes the incompatibility index $e$, defined as a universal relative discrepancy between the internal and external marginal subgroup CATEs, ranging from $0$ to $0.1$ in increments of $0.002$. At each value of $e$, results are based on $5{,}000$ Monte Carlo replications with an internal sample size of $n=500$. Abbreviation: CATE, conditional average treatment effect.}
    \label{fig:prevalence}
\end{figure}

\paragraph{Target Risk Comparison} Figure~\ref{fig:prevalence} presents the relative target risk under prevalence weighting for the fine-grained subgroup CATE vector (defined jointly by $X_1$ and $X_2$), benchmarked against the internal-only unconstrained estimator. Our proposed JS estimator consistently achieves lower target risk than the unconstrained estimator, providing simulation-based confirmation of Theorem~\ref{thm:js}. When the compatibility condition~\ref{compatibility} between the external and internal marginal subgroup CATEs (defined separately by $X_1$ or $X_2$) holds or is only mildly violated, the JS estimator achieves greater efficiency than the EB estimator by more effectively shrinking toward the constrained estimator that incorporates external information, i.e., placing less weight on the unconstrained estimator (Web Figure S1). For example, under full compatibility ($e=0$), the JS estimator attains a relative target risk of $0.62$, compared to $0.78$ for the EB estimator. As compatibility deteriorates, the JS estimator remains stable, whereas the constrained and generalized ridge estimators exhibit marked performance degradation. Results under alternative weighting schemes for the target risk exhibit the same pattern (Web Figure S2), further supporting the risk dominance result in Theorem~\ref{thm:js}.

\begin{figure}
    \centering
    \includegraphics[width=1\linewidth]{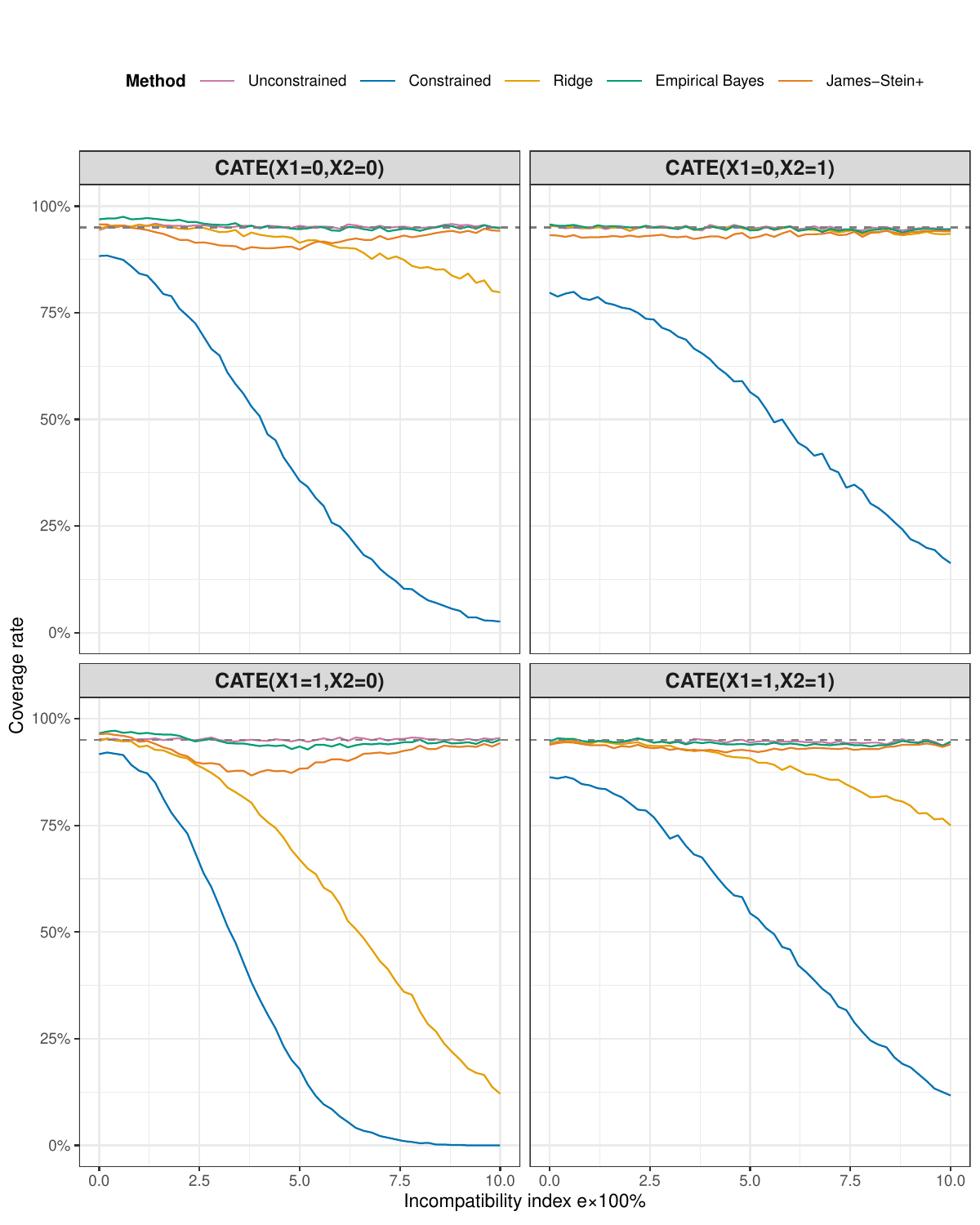}
    \caption{\textbf{Empirical coverage of naive Wald-type 95\% CIs for fine-grained subgroup-specific CATEs across methods.} The horizontal axis denotes the incompatibility index $e$, defined as a universal relative discrepancy between the internal and external marginal subgroup CATEs, ranging from $0$ to $0.1$ in increments of $0.002$. At each value of $e$, results are based on $5{,}000$ Monte Carlo replications with an internal sample size of $n=500$. The horizontal dashed line indicates the nominal 95\% coverage level. Abbreviations: CI, confidence interval; CATE, conditional average treatment effect.}
    \label{fig:simulation}
\end{figure}

\paragraph{Subgroup-Specific CATE Estimation and CI Coverage} In practice, interest often lies in individual fine-grained subgroup CATEs rather than the full CATE vector. We therefore evaluate the four subgroup-specific CATEs jointly defined by $X_1$ and $X_2$ for the proposed JS estimator. Web Figure S3 shows that the proposed analytic variance estimator is slightly conservative relative to the Monte Carlo variance and performs favorably compared with competing estimators (see Web Appendix F.1 for more discussions). Figure~\ref{fig:simulation} reports empirical coverage rates of naive Wald-type $95\%$ CIs (without bias adjustment) for each subgroup-specific CATE, alongside results from alternative CATE estimators. Across subgroups and values of the incompatibility index, coverage for the JS estimator ranges from $86.70\%$ to $96.50\%$, with an average of $92.67\%$ (see also Web Table S3). The naive Wald-type CI exhibits mild undercoverage. 

We note two points. First, the JS estimator does not admit an exact normal sampling distribution, even under normal regression errors (Web Figures S4--9). Second, risk dominance does not necessarily hold at the level of individual subgroup-specific CATEs (Web Figure S10 and Web Appendix F.2), providing empirical support for the conditions in Theorem~\ref{thm:js}. 

\section{Application: Race-by-Sex Subgroup CATE Estimation in a tirzepatide (Zepbound) Obesity Trial}

\begin{figure}
    \centering
    \includegraphics[width=1\linewidth]{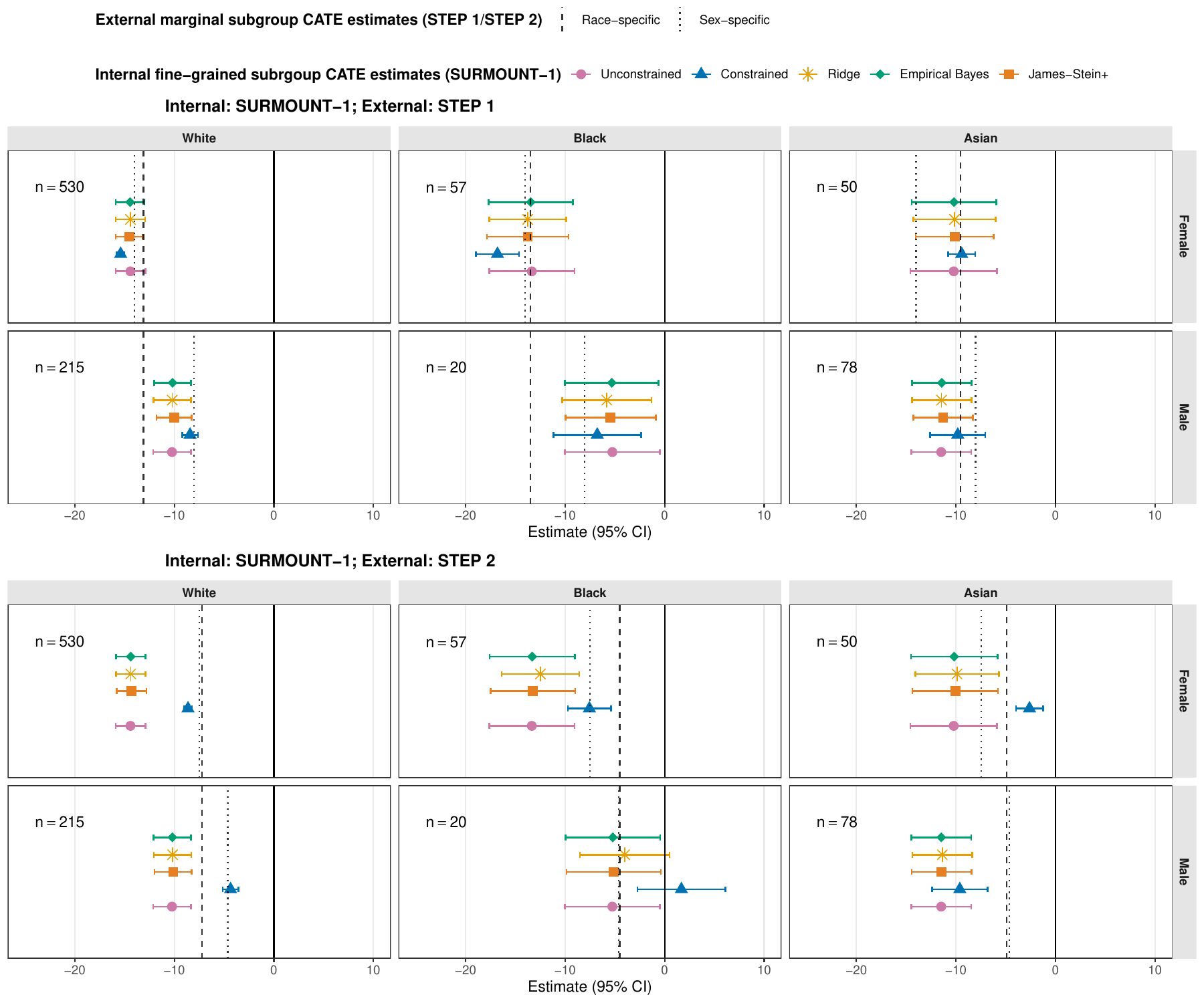}
    \caption{\textbf{Estimates and naive Wald-type 95\% CIs for fine-grained subgroup-specific CATEs of 5 mg tirzepatide on percent change in body weight in SURMOUNT-1 across all methods.} External race-specific and sex-specific CATE estimates for 2.4 mg semaglutide from either STEP~1 or STEP~2 are indicated by black dashed and dotted lines, respectively. The solid black line denotes zero percent change in body weight. Each box reports the internal sample size for the corresponding fine-grained subgroup. Abbreviations: CI, confidence interval; CATE, conditional average treatment effect}
    \label{fig:realdata}
\end{figure}

We apply the proposed JS estimator, along with the alternative estimators described in Section~4, to estimate the race-by-sex subgroup-specific CATEs of 5 mg tirzepatide on percent change in body weight in the SURMOUNT-1 trial \citep{jastreboff2022tirzepatide}. External information is obtained from marginal subgroup CATE estimates defined separately by sex and by race for 2.4 mg semaglutide from STEP~1 and STEP~2 \citep{wilding2021once,davies2021semaglutide}. The publicly available coarsened estimates from the STEP trials are presented in Figure \ref{fig:realdata}, with detailed comparisons across trials provided in Web Table S4. Although the compatibility condition~\ref{compatibility} is not testable with observed data, we expect the coarsened external information from STEP~1 to be more compatible with SURMOUNT-1, as both trials enrolled participants without diabetes. Nonetheless, some incompatibility may persist due to differences between the treatments.

Figure \ref{fig:realdata} displays estimates of the target CATEs for fine-grained subgroups defined jointly by sex and race, along with the corresponding naive Wald-type $95\%$ CIs, across all methods. The JS estimator is constructed using prevalence-weighted target risk for the fine-grained subgroup CATE vector, thereby assigning greater weight to more prevalent subgroups. Results under alternative weighting schemes are reported in Web Figure S11, where we also examine performance under a smaller internal sample by constructing a stratified half-sample of the SURMOUNT-1 data by sex, race, and treatment.

\paragraph{STEP~1} When the coarsened external information is relatively compatible with the internal data (i.e., drawn from STEP~1), the proposed JS estimator effectively shrinks the unconstrained estimator toward the constrained estimator for the target fine-grained subgroup CATEs. In this setting, the weight assigned to the unconstrained estimator is $0.88$ under the proposed JS estimator, compared with $0.96$ for the EB estimator. Using the largest fine-grained subgroup, the female-White subgroup ($n = 530$), as an example, the JS estimate of the subgroup-specific CATE is $-14.53$ [95\% CI: $-15.89, -13.16$], reducing the CI length $91.48\%$ of that of the unconstrained estimate, $-14.41$ [95\% CI: $-15.90, -12.92$]. After borrowing information across sex-specific and race-specific CATEs, the JS estimator indicates a statistically significant difference ($p = 0.04$) between the female-White and female-Asian subgroups ($n = 50$), with the latter estimated as $-10.19$ [95\% CI: $-14.05, -6.21$]. In contrast, the unconstrained estimator based solely on SURMOUNT-1 data does not provide sufficient evidence for such a difference ($p = 0.07$), yielding an estimate of $-10.22$ [95\% CI: $-14.58, -5.86$] for the female-Asian subgroup. Compared with alternative shrinkage methods, the JS estimator exhibits the greatest degree of shrinkage toward the constrained estimator in both point estimation and precision. Specifically, the generalized ridge estimate is $-14.41$ [95\% CI: $-15.89, -12.93$], and the EB estimate is $-14.44$ [95\% CI: $-15.89, -12.99$].

\paragraph{STEP~2} When the coarsened external summary information is substantially incompatible with the internal data (i.e., drawn from STEP~2), the proposed JS estimator exhibits robustness by placing nearly all weight on the unconstrained estimator, with a weight of $0.98$ in this setting. Using the same example as in STEP~1, the JS estimate of the subgroup-specific CATE for the female-White subgroup ($n = 530$) is $-14.30$ [95\% CI: $-15.79, -12.81$], closely aligning the unconstrained estimate of $-14.41$ [95\% CI: $-15.90, -12.92$], rather than the constrained estimate of $-8.63$ [95\% CI: $-9.01, -8.24$].

\section{Discussion}

The proposed JS method contributes to the existing literature by leveraging the coarsened subgroup estimates from an external RCT to improve the estimation of subgroup-specific CATEs within an internal RCT. We show theoretically that the proposed estimator uniformly outperforms the internal-only unconstrained estimator under a target quadratic loss defined on the fine-grained subgroup CATE vector, regardless of the degree of compatibility between marginal subgroup CATEs across trials. Simulation studies further demonstrate that the method achieves effective shrinkage when the coarsened external information is set to be compatible or only mildly incompatible with the internal counterparts, while maintaining robust performance even under substantial incompatibility. In a real data application estimating race-by-sex subgroup CATEs for once-weekly 5 mg tirzepatide in SURMOUNT-1, the proposed method reveals greater treatment heterogeneity. For example, both the proposed method and the internal-only analysis suggest a greater percentage weight loss among female-White subgroups compared with female-Asian subgroups. However, by borrowing external information from the relatively compatible semaglutide trial STEP~1, the proposed method identifies a statistically significant difference between these subgroups ($p = 0.04$), whereas the internal-only analysis does not ($p = 0.07$).

Building on the proposed variance estimator, we construct a naive Wald-type CI for the JS estimator without bias adjustment or length reduction. The average empirical coverage rate is $92.67\%$, falling below the nominal level of $95\%$. Undercoverage of this kind has also been documented in the JS literature. For example, in a related JS-type method for stratified CATEs, \cite{rosenman2023combining} reported undercoverage for the bootstrap-based CI of \cite{laird1987empirical} relative to the nominal EB coverage level of $90\%$, with mean coverage ranging from $88\%$ to $97\%$ across settings and minimum coverage as low as $53\%$. While our work makes an important initial effort toward frequentist coverage for JS-type causal estimators, future work is needed to develop CIs with improved frequentist coverage accuracy and shorter length \citep{casella2012shrinkage}.

\backmatter

\section*{Acknowledgements}

This publication is based on research using data from data contributors Eli Lilly that has been made available through Vivli, Inc. Vivli has not contributed to or approved, and Vivli and Eli Lilly are not in any way responsible for, the contents of this publication. 

\section*{Supplementary Materials}

Web Appendices, Tables, and Figures referenced in Sections 3, 4, 5, and 6 are available at the Biometrics website on Wiley Online Library.
\vspace*{-8pt}

\bibliographystyle{biom} 
\bibliography{main}

@article{rubin1974estimating,
  title={Estimating causal effects of treatments in randomized and nonrandomized studies.},
  author={Rubin, Donald B},
  journal={Journal of educational Psychology},
  volume={66},
  number={5},
  pages={688},
  year={1974},
  publisher={American Psychological Association}
}

@article{rubin1980randomization,
  title={Randomization analysis of experimental data: The Fisher randomization test comment},
  author={Rubin, Donald B},
  journal={Journal of the American statistical association},
  volume={75},
  number={371},
  pages={591--593},
  year={1980},
  publisher={JSTOR}
}

@article{schochet2010regression,
  title={Is regression adjustment supported by the Neyman model for causal inference?},
  author={Schochet, Peter Z},
  journal={Journal of Statistical Planning and Inference},
  volume={140},
  number={1},
  pages={246--259},
  year={2010},
  publisher={Elsevier}
}

@article{yang2001efficiency,
  title={Efficiency study of estimators for a treatment effect in a pretest--posttest trial},
  author={Yang, Li and Tsiatis, Anastasios A},
  journal={The American Statistician},
  volume={55},
  number={4},
  pages={314--321},
  year={2001},
  publisher={Taylor \& Francis}
}

@article{lin2013agnostic,
  title={Agnostic notes on regression adjustments to experimental data: Reexamining Freedman’s critique},
  author={Lin, Winston},
  journal={The Annals of Applied Statistics},
  volume={7},
  number={1},
  pages={295--318},
  year={2013},
  publisher={Institute of Mathematical Statistics}
}

@article{white1980heteroskedasticity,
  title={A heteroskedasticity-consistent covariance matrix estimator and a direct test for heteroskedasticity},
  author={White, Halbert},
  journal={Econometrica: journal of the Econometric Society},
  pages={817--838},
  year={1980},
  publisher={JSTOR}
}

@article{samii2012equivalencies,
  title={On equivalencies between design-based and regression-based variance estimators for randomized experiments},
  author={Samii, Cyrus and Aronow, Peter M},
  journal={Statistics \& Probability Letters},
  volume={82},
  number={2},
  pages={365--370},
  year={2012},
  publisher={Elsevier}
}

@article{jastreboff2022tirzepatide,
  title={Tirzepatide once weekly for the treatment of obesity},
  author={Jastreboff, Ania M and Aronne, Louis J and Ahmad, Nadia N and Wharton, Sean and Connery, Lisa and Alves, Breno and Kiyosue, Arihiro and Zhang, Shuyu and Liu, Bing and Bunck, Mathijs C and others},
  journal={New England Journal of Medicine},
  volume={387},
  number={3},
  pages={205--216},
  year={2022},
  publisher={Mass Medical Soc}
}

@article{wilding2021once,
  title={Once-weekly semaglutide in adults with overweight or obesity},
  author={Wilding, John PH and Batterham, Rachel L and Calanna, Salvatore and Davies, Melanie and Van Gaal, Luc F and Lingvay, Ildiko and McGowan, Barbara M and Rosenstock, Julio and Tran, Marie TD and Wadden, Thomas A and others},
  journal={New England Journal of Medicine},
  volume={384},
  number={11},
  pages={989--1002},
  year={2021},
  publisher={Mass Medical Soc}
}

@article{davies2021semaglutide,
  title={Semaglutide 2{\textperiodcentered} 4 mg once a week in adults with overweight or obesity, and type 2 diabetes (STEP 2): a randomised, double-blind, double-dummy, placebo-controlled, phase 3 trial},
  author={Davies, Melanie and F{\ae}rch, Louise and Jeppesen, Ole K and Pakseresht, Arash and Pedersen, Sue D and Perreault, Leigh and Rosenstock, Julio and Shimomura, Iichiro and Viljoen, Adie and Wadden, Thomas A and others},
  journal={The Lancet},
  volume={397},
  number={10278},
  pages={971--984},
  year={2021},
  publisher={Elsevier}
}

@article{konwar2022efficacy,
  title={Efficacy and Safety of Liraglutide 3.0 mg in Patients with Overweight and Obese with or without Diabetes: A Systematic Review and Meta-Analysis},
  author={Konwar, Mahanjit and Bose, Debdipta and Jaiswal, Sanjeet Kumar and Maurya, Mitesh Kumar and Ravi, Renju},
  journal={International journal of clinical practice},
  volume={2022},
  number={1},
  pages={1201977},
  year={2022},
  publisher={Wiley Online Library}
}

@article{alsaqaaby2024sex,
  title={Sex, race, and BMI in clinical trials of medications for obesity over the past three decades: a systematic review},
  author={Alsaqaaby, Moath S and Cooney, Sarah and le Roux, Carel W and Pournaras, Dimitri J},
  journal={The Lancet Diabetes \& Endocrinology},
  year={2024},
  publisher={Elsevier}
}

@article{kent2010assessing,
  title={Assessing and reporting heterogeneity in treatment effects in clinical trials: a proposal},
  author={Kent, David M and Rothwell, Peter M and Ioannidis, John PA and Altman, Doug G and Hayward, Rodney A},
  journal={Trials},
  volume={11},
  pages={1--11},
  year={2010},
  publisher={Springer}
}

@article{wang2007statistics,
  title={Statistics in medicine—reporting of subgroup analyses in clinical trials},
  author={Wang, Rui and Lagakos, Stephen W and Ware, James H and Hunter, David J and Drazen, Jeffrey M},
  journal={New England Journal of Medicine},
  volume={357},
  number={21},
  pages={2189--2194},
  year={2007},
  publisher={Mass Medical Soc}
}

@article{tierney2020comparison,
  title={Comparison of aggregate and individual participant data approaches to meta-analysis of randomised trials: an observational study},
  author={Tierney, Jayne F and Fisher, David J and Burdett, Sarah and Stewart, Lesley A and Parmar, Mahesh KB},
  journal={PLoS medicine},
  volume={17},
  number={1},
  pages={e1003019},
  year={2020},
  publisher={Public Library of Science San Francisco, CA USA}
}

@article{shi2023data,
  title={Data integration in causal inference},
  author={Shi, Xu and Pan, Ziyang and Miao, Wang},
  journal={Wiley Interdisciplinary Reviews: Computational Statistics},
  volume={15},
  number={1},
  pages={e1581},
  year={2023},
  publisher={Wiley Online Library}
}

@article{brantner2023methods,
  title={Methods for integrating trials and non-experimental data to examine treatment effect heterogeneity},
  author={Brantner, Carly Lupton and Chang, Ting-Hsuan and Nguyen, Trang Quynh and Hong, Hwanhee and Di Stefano, Leon and Stuart, Elizabeth A},
  journal={Statistical science: a review journal of the Institute of Mathematical Statistics},
  volume={38},
  number={4},
  pages={640},
  year={2023}
}

@article{kovalchik2013aggregate,
  title={Aggregate-data estimation of an individual patient data linear random effects meta-analysis with a patient covariate-treatment interaction term},
  author={Kovalchik, Stephanie A},
  journal={Biostatistics},
  volume={14},
  number={2},
  pages={273--283},
  year={2013},
  publisher={Oxford University Press}
}

@article{godolphin2023estimating,
  title={Estimating interactions and subgroup-specific treatment effects in meta-analysis without aggregation bias: A within-trial framework},
  author={Godolphin, Peter J and White, Ian R and Tierney, Jayne F and Fisher, David J},
  journal={Research Synthesis Methods},
  volume={14},
  number={1},
  pages={68--78},
  year={2023},
  publisher={Wiley Online Library}
}

@article{van2023conditional,
  title={Conditional average treatment effect estimation with marginally constrained models},
  author={Van Amsterdam, Wouter AC and Ranganath, Rajesh},
  journal={Journal of Causal Inference},
  volume={11},
  number={1},
  pages={20220027},
  year={2023},
  publisher={De Gruyter}
}

@inproceedings{JamesStein1961,
  title     = {Estimation with Quadratic Loss},
  author    = {James, W. and Stein, C.},
  booktitle = {Proceedings of the Fourth Berkeley Symposium on Mathematical Statistics and Probability, Volume 1: Contributions to the Theory of Statistics},
  pages     = {361--379},
  year      = {1961},
  publisher = {University of California Press}
}

@article{green1991james,
  title={A James-Stein type estimator for combining unbiased and possibly biased estimators},
  author={Green, Edwin J and Strawderman, William E},
  journal={Journal of the American Statistical Association},
  volume={86},
  number={416},
  pages={1001--1006},
  year={1991},
  publisher={Taylor \& Francis}
}

@article{han2024improving,
  title={Improving prediction of linear regression models by integrating external information from heterogeneous populations: James--Stein estimators},
  author={Han, Peisong and Li, Haoyue and Park, Sung Kyun and Mukherjee, Bhramar and Taylor, Jeremy MG},
  journal={Biometrics},
  volume={80},
  number={3},
  pages={ujae072},
  year={2024},
  publisher={Oxford University Press}
}

@article{jennrich1986much,
  title={How much does Stein estimation help in multiple linear regression?},
  author={Jennrich, Robert I and Oman, Samuel D},
  journal={Technometrics},
  volume={28},
  number={2},
  pages={113--121},
  year={1986},
  publisher={Taylor \& Francis}
}

@article{rosenman2023combining,
  title={Combining observational and experimental datasets using shrinkage estimators},
  author={Rosenman, Evan TR and Basse, Guillaume and Owen, Art B and Baiocchi, Mike},
  journal={Biometrics},
  volume={79},
  number={4},
  pages={2961--2973},
  year={2023},
  publisher={Wiley Online Library}
}

@article{hansen2016efficient,
  title={Efficient shrinkage in parametric models},
  author={Hansen, Bruce E},
  journal={Journal of Econometrics},
  volume={190},
  number={1},
  pages={115--132},
  year={2016},
  publisher={Elsevier}
}

@article{baranchik1970family,
  title={A family of minimax estimators of the mean of a multivariate normal distribution},
  author={Baranchik, Alvin J},
  journal={The Annals of Mathematical Statistics},
  pages={642--645},
  year={1970},
  publisher={JSTOR}
}

@article{casella2012shrinkage,
  title={Shrinkage confidence procedures},
  author={Casella, George and Hwang, JT Gene},
  journal={Statistical Science},
  pages={51--60},
  year={2012},
  publisher={JSTOR}
}

@article{laird1987empirical,
  title={Empirical Bayes confidence intervals based on bootstrap samples},
  author={Laird, Nan M and Louis, Thomas A},
  journal={Journal of the American Statistical Association},
  volume={82},
  number={399},
  pages={739--750},
  year={1987},
  publisher={Taylor \& Francis}
}

@article{chen2009shrinkage,
  title={Shrinkage estimators for robust and efficient inference in haplotype-based case-control studies},
  author={Chen, Yi-Hau and Chatterjee, Nilanjan and Carroll, Raymond J},
  journal={Journal of the American Statistical Association},
  volume={104},
  number={485},
  pages={220--233},
  year={2009},
  publisher={Taylor \& Francis}
}

@article{mukherjee2008exploiting,
  title={Exploiting gene-environment independence for analysis of case--control studies: an empirical Bayes-type shrinkage estimator to trade-off between bias and efficiency},
  author={Mukherjee, Bhramar and Chatterjee, Nilanjan},
  journal={Biometrics},
  volume={64},
  number={3},
  pages={685--694},
  year={2008},
  publisher={Oxford University Press}
}

@article{colnet2024causal,
  title={Causal inference methods for combining randomized trials and observational studies: a review},
  author={Colnet, B{\'e}n{\'e}dicte and Mayer, Imke and Chen, Guanhua and Dieng, Awa and Li, Ruohong and Varoquaux, Ga{\"e}l and Vert, Jean-Philippe and Josse, Julie and Yang, Shu},
  journal={Statistical science},
  volume={39},
  number={1},
  pages={165--191},
  year={2024},
  publisher={Institute of Mathematical Statistics}
}

@article{qin1994empirical,
  title={Empirical likelihood and general estimating equations},
  author={Qin, Jin and Lawless, Jerry},
  journal={the Annals of Statistics},
  volume={22},
  number={1},
  pages={300--325},
  year={1994},
  publisher={Institute of Mathematical Statistics}
}

@article{stein1981estimation,
  title={Estimation of the mean of a multivariate normal distribution},
  author={Stein, Charles M},
  journal={The annals of Statistics},
  pages={1135--1151},
  year={1981},
  publisher={JSTOR}
}

@article{yi1991estimating,
  title={Estimating the variability of the Stein estimator by bootstrap},
  author={Yi, Gang},
  journal={Economics Letters},
  volume={37},
  number={3},
  pages={293--298},
  year={1991},
  publisher={Elsevier}
}

@article{armstrong2022robust,
  title={Robust empirical Bayes confidence intervals},
  author={Armstrong, Timothy B and Koles{\'a}r, Michal and Plagborg-M{\o}ller, Mikkel},
  journal={Econometrica},
  volume={90},
  number={6},
  pages={2567--2602},
  year={2022},
  publisher={Wiley Online Library}
}

@article{boss2025mediation,
  title={Mediation with External Summary Statistic Information},
  author={Boss, Jonathan and Hao, Wei and Cathey, Amber and Welch, Barrett M and Ferguson, Kelly K and Meeker, John D and Zhou, Xiang and Kang, Jian and Mukherjee, Bhramar},
  journal={Biostatistics},
  volume={26},
  number={1},
  pages={kxaf020},
  year={2025},
  publisher={Oxford University Press}
}

\label{lastpage}

\end{document}